%% file: main.tex
\documentclass[conference]{IEEEtran}
\IEEEoverridecommandlockouts
\usepackage{cite}
\usepackage{amsmath,amssymb,amsfonts}
\usepackage{algorithm}

\usepackage{graphicx}
\usepackage{textcomp}
\usepackage{xcolor}
\usepackage{subcaption}
\usepackage{multirow}
\usepackage{lipsum}
\usepackage{mathtools}
\usepackage{cuted}
\usepackage{stfloats}
\usepackage{amsthm}

\usepackage{algpseudocode}%
\usepackage{balance}

\usepackage{tabularx}
\makeatletter
\newcommand{\multiline}[1]{%
  \begin{tabularx}{\dimexpr\linewidth-\ALG@thistlm}[t]{@{}X@{}}
    #1
  \end{tabularx}
}
\makeatother

\usepackage[font=normal]{caption}
\usepackage[binary-units=true]{siunitx} 

\usepackage{enumitem}

\allowdisplaybreaks

\begin{document}

\title{
Deep Reinforcement Learning-Based Adaptive IRS Control with Limited Feedback Codebooks
}
\author{\IEEEauthorblockN{Junghoon Kim\IEEEauthorrefmark{1}, Seyyedali Hosseinalipour\IEEEauthorrefmark{1}, Andrew C. Marcum\IEEEauthorrefmark{2},
Taejoon Kim\IEEEauthorrefmark{3}, \\
David J. Love\IEEEauthorrefmark{1} and Christopher G. Brinton\IEEEauthorrefmark{1}}
\IEEEauthorblockA{\IEEEauthorrefmark{1}\selectfont{Electrical and Computer Engineering, Purdue University, West Lafayette, IN, USA}}
\IEEEauthorblockA{\IEEEauthorrefmark{2}Raytheon BBN Technologies, Cambridge, MA, USA}
\IEEEauthorblockA{\IEEEauthorrefmark{3}Electrical Engineering and Computer Science, University of Kansas, Lawrence, KS, USA}
\IEEEauthorblockA{\IEEEauthorrefmark{1}\{kim3220, hosseina, djlove, cgb\}@purdue.edu,  \IEEEauthorrefmark{2}andrew.marcum@raytheon.com,
\IEEEauthorrefmark{3}taejoonkim@ku.edu}
\thanks{A more comprehensive version of this paper is under review in IEEE Transactions on Wireless Communications~\cite{kim2021learning}.
C. G. Brinton was supported in part by the
National Spectrum Consortium (NSC) under Grant W15QKN-15-9-1004 and Office of Naval Research (ONR) under Grant N00014-21-1-2472.
T. Kim was supported in part by the National Science Foundation (NSF) under Grants CNS1955561.
}
}

\maketitle

\begin{abstract}

\input{abstract}

\end{abstract}

\input{intro}
\input{system}

\input{formulation}
\input{method}

\input{sim}

\input{conc}






\bibliographystyle{IEEEtran}
\balance
\bibliography{ref}

\end{document}

%% file: abstract.tex
Intelligent reflecting surfaces (IRS) consist of configurable meta-atoms, which can alter the wireless propagation environment through design of their reflection coefficients.
We consider adaptive 
IRS 
control 
in the practical setting where
(i) the IRS reflection coefficients are attained by adjusting \textit{tunable elements} embedded in the meta-atoms,
(ii) the IRS reflection coefficients are affected by the \textit{incident angles} of the incoming signals,
(iii) the IRS is deployed in multi-path, time-varying channels, and
(iv) the feedback link from the base station (BS) to the IRS has a low data rate. 
Conventional optimization-based IRS control protocols, which
rely on channel estimation and conveying the optimized variables to the IRS, are not practical in this setting due to the difficulty of channel estimation and the low data rate of the feedback channel. 
To address these challenges, we develop a novel adaptive codebook-based limited feedback protocol to control the IRS. 
We propose two solutions for adaptive IRS codebook design:
(i)
\textit{random adjacency (RA)}, which utilizes  correlations across the channel realizations, and (ii) \textit{deep neural network policy-based IRS control (DPIC)}, which is based on a deep reinforcement learning.
Numerical evaluations show that 
the data rate and average data rate over one coherence~time
are improved substantially by the proposed schemes.

%% file: intro.tex
\section{Introduction}
\label{sec:intro}

The intelligent reflecting surface (IRS) is a technology for 6G-and-beyond~\cite{zhang2020prospective,hosseinalipour2020federated}. 
An IRS  is a software-controlled meta-surface, consisting of configurable meta-atoms with flexible reflection coefficients.
By fine-tuning these meta-atoms, the IRS can change the wireless propagation environment,
resulting in power savings, throughput increase,~etc.
Compared to a traditional antenna array with radio frequency (RF) chains for
active relaying/beamforming,
an IRS is made of low cost meta-surfaces,
which consume low energy for tuning~\cite{wu2019towards}.
These benefits have motivated research on utilizing IRS
in communications/signal processing literature. 
We aim to address three
shortcomings of the current art,
as discussed below.

\subsection{Shortcomings of Existing Works and Motivations}\label{subsIntro1}
\subsubsection{Dependency between Meta-Atoms' Reflection Phase Shift and Attenuation}\label{intro1} 
Much of the existing work on IRS reflection coefficient design for communications has treated
the reflection phase and attenuation of each meta atom 
independently~\cite{yang2020intelligent}.
In reality, 
the phase shift and attenuation are \textit{interdependent}
because the reflection behavior is determined by adjusting the \textit{tunable elements} inside the meta-atoms, i.e., their controllable capacitance,
as revealed in physics literature~\cite{shao2021electrically}. 
This interdependency has only been considered in a few works in the communications area~\cite{abeywickrama2020intelligent}.

\subsubsection{Dependency between Meta-atoms' Reflection Coefficient and Incident Angle of Incoming Signals}\label{intro2} Another practical consideration neglected in existing works is the dependency between the IRS reflection behavior and the \textit{incident angles} of the electromagnetic (EM) waves~\cite{pei2021ris,chen2020angle}.
In~\cite{chen2020angle}, the authors propose an angle-dependent reflection coefficient model for each meta-atom using an equivalent circuit model.
To the best of our knowledge, the angle-dependent property of the IRS reflection coefficient has not been 
incorporated into 
uplink/downlink signal transmission models for wireless communication systems.


\subsubsection{Low Overhead Feedback Channel}
The configuration of IRS meta-atoms is usually controlled via reception of some information from the base station (BS) through a \textit{feedback} link.
This link typically has a low data rate since its channel state information is unknown to the BS~\cite{wu2019towards}.
To reduce the feedback overhead for IRS control, some recent works have considered  
codebook structures~\cite{pei2021ris,kim2021multi}, where
the codebook refers to a set of IRS reflection coefficients.
In these works,
the BS feeds back a specific codeword index to the IRS, using which the IRS recovers the desired reflection coefficients from the codebook. 
In doing so, these works
directly design the IRS {\it reflection coefficients} and thus do not consider the practical IRS reflection behavior mentioned  in~Sec.~\ref{intro1},~\ref{intro2}.

\subsection{Overview of Methodology and Contributions}

We propose a new methodology for IRS control that explicitly considers
the above three practical design aspects.
%
In doing so, we consider that the IRS is deployed in realistic multi-path, time-varying channels. 
These considerations render  current optimization-based methods for IRS control~\cite{yang2020intelligent,abeywickrama2020intelligent}, which rely on channel estimation, impractical:
it is difficult to measure the incident angles of incoming signals at the IRS since the IRS typically
does not have active sensors.





Specifically,
we propose a novel \textit{adaptive codebook-based limited feedback protocol.}
We directly design the meta-atom {\it capacitance values}, instead of their reflection coefficients as in existing methods. 
With the codebook  as a set of capacitance values for the meta-atoms, we develop two \textit{adaptive} codebook design methods:
(i) \textit{random adjacency (RA)}
and (ii) \textit{deep neural network  policy-based IRS control (DPIC)}.
These approaches only require the end-to-end channel from the user equipment (UE) to the BS, which can be readily measured in real-time.

%% file: system.tex
\section{System Model for IRS-assisted Communications}
\label{sec:signal}
\vspace{-1mm}
We begin by formalizing IRS meta-atom reflection behavior
in Sec.~\ref{ssec:IRS}.
Then, we describe the signal model of IRS-assisted uplink communications in Sec.~\ref{ssec:formulation}.

\begin{figure}[t]
  \includegraphics[width=\linewidth]{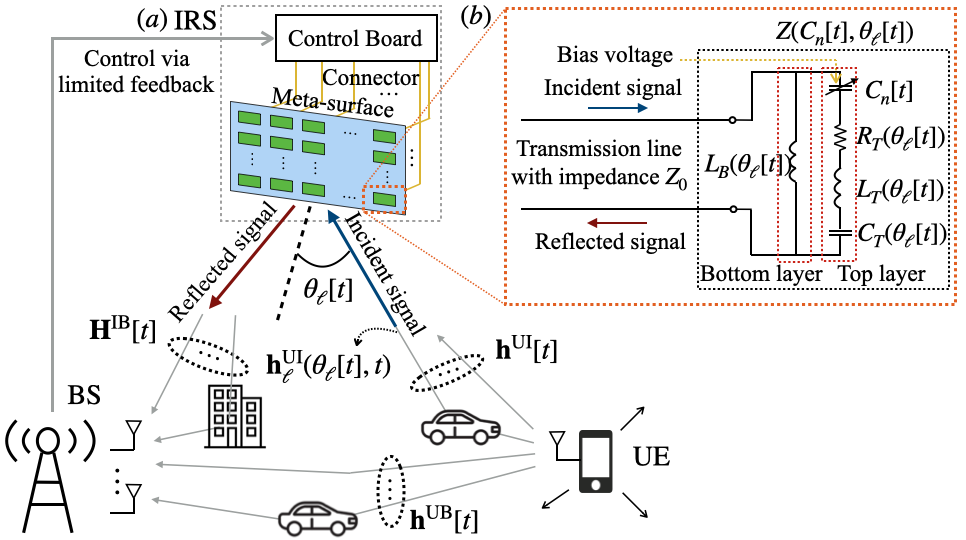}
  \centering
  \caption{An uplink point-to-point communication system consisting of a UE, IRS, and BS, where
  the IRS is controlled by the BS via a limited feedback link.
  (a) 
  IRS as two interconnected systems: meta-surface and control board.
  (b) Equivalent circuit model 
  of each meta-atom.}
  \label{fig:system_merged}
\end{figure}

\subsection{Reflection Behavior of IRS Meta-atoms}
\label{ssec:IRS}

An IRS is composed of two interconnected systems as shown in Fig. \ref{fig:system_merged}(a): a meta-surface 
and a control board. 
A meta-surface is 
an ultra-thin sheet composed of 
periodic sub-wavelength metal/dielectric structures, i.e., meta-atoms.
Each meta-atom generally contains a semiconductor device, i.e., the tunable element, such as a positive-intrinsic-negative (PIN) diode and a variable capacitor (varactor) \cite{shao2021electrically}.
%
A control board, e.g., a field programmable gate array (FPGA)~\cite{abadal2020programmable}, adjusts the bias voltage applied to the semiconductor in each meta-atom and changes its capacitance.
Given a range of potential bias voltage values,
the capacitance $C_n[t]$ at meta-atom $n$ at time $t$ satisfies
\begin{equation}
    C_{\min} \le C_n[t] \le C_{\max},
    \label{eq:Crange}
\end{equation}
where $C_{\min}$ and $C_{\max}$ 
depend on the semiconductor used.

Through tuning the capacitance of the meta-atoms, 
their impedance
can be adjusted. 
However, the impedance is also dependent on the incident angle of the incoming EM wave~\cite{chen2020angle,pei2021ris}.
Both of these factors should be considered in IRS reflection behavior design.
As an example, we provide 
the impedance  and reflection coefficient of a meta-atom equipped with a \textit{varactor} using its equivalent circuit model~\cite{chen2020angle}
depicted
in Fig. \ref{fig:system_merged}(b).
Let $\theta_\ell[t]$ denote the incident angle of the $\ell$-th channel path to the IRS.\footnote{
We are discussing the angle-dependent reflection model provided in \cite{chen2020angle} where only azimuth coordinates of the incident angle are considered. 
}
Under a far-field assumption where $\theta_\ell[t]$ is the same across all the meta-atoms, the impedance of
 meta-atom $n$ can be described
as~\cite{chen2020angle} 

\vspace{-5mm}
\small
\begin{align}
    &\hspace{3mm} Z(C_n[t], \theta_\ell[t]) =
    \label{eq:Z}    
    \\
    & \frac{
    j\omega L_B(\theta_\ell[t]) \big( R_T(\theta_\ell[t]) + j\omega L_T(\theta_\ell[t]) +
    \frac{1}{j\omega C_T(\theta_\ell[t])}   + \frac{1}{j\omega C_n[t]}
    \big)}
    %
    {
    j\omega L_B(\theta_\ell[t]) + R_T(\theta_\ell[t]) + j\omega L_T(\theta_\ell[t]) + \frac{1}{j \omega C_T(\theta_\ell[t])}   + \frac{1}{j \omega C_n[t]} 
    },
    \nonumber
    \vspace{-1.5mm}
\end{align}
\normalsize
where {$L_T(\cdot)$}, {$C_T(\cdot)$}, and { $R_T (\cdot)$} are the inductance, capacitance, and resistance of the top circuit layer in Fig. \ref{fig:system_merged}(b), respectively,
 { $L_B(\cdot)$} is the bottom layer inductance,
 $C_n[t]$ is the variable capacitance, and $\omega$ is the angular  frequency of the EM waves.

Considering the impedance discontinuity between the free space impedance 
{$Z_0 \approx  \SI{376.73}{\ohm}$}
and the meta-atom impedance { $Z(C_n[t], \theta_\ell[t])$}, the reflection coefficient\footnote{
We assume a narrowband system with a few tens of MHz in bandwidth.
Then, we can approximate the reflection coefficients as constant
across~$\omega$~\cite{chen2020angle, pei2021ris}.}
%
of meta-atom $n$ is~\cite{chen2020angle}
\begin{equation}
    \Gamma(C_n[t], \theta_\ell[t]) = \frac{Z(C_n[t], \theta_\ell[t]) - Z_0}
    {Z(C_n[t], \theta_\ell[t]) + Z_0}.
    \label{eq:reflectioncoeff}
\end{equation}

The expressions in \eqref{eq:Z}\&\eqref{eq:reflectioncoeff} reveal
two practical considerations for tuning the meta-atoms.
First, the reflection attenuation $|\Gamma(C_n[t], \theta_\ell[t])|$ and phase shift $\angle \Gamma(C_n[t], \theta_\ell[t])$ are jointly controlled by 
the variable capacitance $C_n[t]$, as also reported in~\cite{abeywickrama2020intelligent}.
Thus, it is beneficial to design the variable capacitance instead of the reflection coefficient since some combinations of attenuation and phase shifts may not be feasible.
Second, the reflection coefficient is a function of the incident angle $\theta_\ell[t]$, posing new challenges for applications of IRS in 
multi-path and time-varying channels, which will be discussed in Sec. \ref{ssec:problem}. 
While observed  in~\cite{chen2020angle,pei2021ris}, 
this dependency has not yet been
incorporated in the canonical signal model for IRS-assisted communications.
In this paper, we incorporate these practical considerations into our signal model and~methodology.

\subsection{Signal Model for IRS-assisted Uplink Communications}
\label{ssec:formulation}



We consider IRS-assisted uplink communications with a UE, a BS, and an IRS, shown in Fig. \ref{fig:system_merged}. 
The UE posesses a single antenna, while the BS has $N_{\rm BS}$ antennas.
%
We assume a block fading channel model with time index  $t=0,1,...$, where channels are constant during each block.
Let $N_{\rm IRS}$ denote the number of IRS meta-atoms.
We define the {\it capacitance vector} across the meta-atoms at time $t$ 
as
\begin{equation}
    {\bf c}[t] = \big[ C_{1}[t], ...,  C_{N_{\rm IRS}}[t] \big] \in \mathbb{R}^{N_{\rm IRS} }.
    \label{eq:capvec}
\end{equation}
We also formulate the {\it reflection coefficient matrix} ${\boldsymbol \Phi}( {\bf c}[t] , \theta_\ell[t] ) \in \mathbb{C}^{N_{\rm IRS} \times N_{\rm IRS} }$ across the IRS meta-atoms~as
\begin{equation}
     {\boldsymbol \Phi}( {\bf c}[t] , \theta_\ell[t] ) \negmedspace = \negmedspace {\rm diag} \big( 
    \Gamma(C_1[t], \theta_\ell[t]), 
    ...,
     \Gamma(C_{N_{\rm IRS}}[t], \theta_\ell[t]) 
      \big), 
     \label{eq:reflection_matrix}
\end{equation}
where the $n$-th diagonal $\Gamma(C_n[t], \theta_\ell[t])$ is the reflection coefficient at meta-atom $n$, $n \in \{1,..., N_{\rm IRS}\}$,
given the incident angle $\theta_\ell[t]$.

We consider multi-path single tap channels and adopt a geometric channel model representation~\cite{tse2005fundamentals}.
The channel from the UE to the IRS 
is described as 
\begin{equation}
    {\bf h}^{\rm UI}[t]  =  \sum_{\ell=1}^{L[t]} {\bf h}^{\rm UI}_\ell(\theta_\ell[t], t) \in  \mathbb{C}^{N_{\rm IRS} \times 1},
\end{equation}
in which
${\bf h}^{\rm UI}_\ell(\theta_\ell[t], t)$ is the $\ell$-th path channel with incident angle $\theta_\ell[t]$ and $L[t]$ is the number of paths. 
The received signal at the BS at time $t$ is given by
\begin{equation}
    {\bf y}[t] =
    {\bf h}_{\rm eff}({\bf c}[t],t)
     \sqrt{P} x[t] + {\bf n}[t] \in \mathbb{C}^{N_{\rm BS} \times 1},
    \label{eq:signal_model}
\end{equation}
where 
$P\geq 0$ is the transmit power and $x[t] \in \mathbb{C}$ is the transmit symbol of the UE with $\mathbb{E}[|x[t]|^2]=1$.
The noise vector ${\bf n}[t]$ follows the
complex Gaussian distribution 
$\mathcal{CN} ( {\bf 0}, \sigma^2 {\bf I} )$,
where ${\bf I}$ denotes the identity matrix and $\sigma^2$ is the variance. In \eqref{eq:signal_model},
the \textit{end-to-end compound channel} ${\bf h}_{\rm eff}({\bf c}[t],t) \in \mathbb{C}^{N_{\rm BS} \times 1}$ is 
\begin{align}
    {\bf h}_{\rm eff}({\bf c}[t],t) &= {\bf h}^{\rm UB}[t] \nonumber
    \\ 
    & \hspace{.7cm} 
    + {\bf H}^{\rm IB}[t] 
    \sum_{\ell =1}^{L[t]} {\boldsymbol \Phi}( {\bf c}[t] , \theta_\ell[t] ) {\bf h}^{\rm UI}_\ell(\theta_\ell[t], t),
    \label{eq:effchan} 
\end{align}
where ${\bf h}^{\rm {UB}}[t] \in \mathbb{C}^{N_{\rm BS} \times 1}$ is the direct channel from the UE to the BS
and ${\bf H}^{\rm IB}[t] \in \mathbb{C}^{N_{\rm BS} \times N_{\rm IRS} }$ is the channel from the IRS to the BS.
${\bf h}_{\rm eff}({\bf c}[t],t)$  
encapsulates all the channels (i.e.,  ${\bf h}^{\rm {UB}}[t]$, ${\bf H}^{\rm IB}[t]$, and ${\bf h}^{\rm UI}[t]$) and the IRS configuration~(i.e.,~${\bf c}[t]$). 
%


%% file: formulation.tex
\section{Problem Formulation and Limited Feedback Protocol}
\label{sec:optimization}

We first formulate the data rate maximization problem for IRS control and discuss the challenges associated with solving it in Sec.~\ref{ssec:problem}. To address the challenges, we propose an adaptive codebook-based limited feedback protocol 
in Sec.~\ref{ssec:protocol}.

\subsection{Problem Formulation and Challenges}
\label{ssec:problem}
%
We formulate the data rate maximization at time~$t$ as
\begin{align}
&  \underset{ {\bf c}[t] }{\text{maximize}} & & \hspace{-1mm} 
 R({\bf c}[t],t) =
W \log_2 
\bigg( 
1+  \frac{ P \| {\bf h}_{\rm eff}( {\bf c}[t],t) \|_2^2} {\sigma^2}
\bigg) 
\hspace{-0.8mm}
\label{eq:obj:rate}
\\
& \text{subject to}
& &  
C_{\min} \le {C}_n[t] \le C_{\max}, \; n=1,..., N_{\rm IRS},
\label{eq:con:rate}
\end{align}
where $W$ is the channel bandwidth (in Hz).
%
In other words, the objective is to adapt ${\bf c}[t]$ based on the time-varying channels.

Operationally, we aim for \eqref{eq:obj:rate}-\eqref{eq:con:rate} to be solved at the BS 
since the BS can obtain measurements 
and has abundant computing~resources.
The BS would then generate feedback information for the IRS.
However, 
solving \eqref{eq:obj:rate}-\eqref{eq:con:rate} 
presents 
three key challenges.
First, conventional optimization-based methods, relying on channel estimations, cannot be applied: 
the channel estimation requires the IRS reflection coefficients, which 
depend on the incident angles of incoming signals that cannot readily be measured (the IRS has no active sensors).
Second, adaptive control of ${\bf c}[t]$ is necessary for efficient IRS operation in time-varying channels, which requires periodic information acquisition from the BS.
Related to this is the third challenge: the feedback link has a low data rate~\cite{wu2019towards}.
%
%
The main contribution of our work is developing methodology to jointly address these challenges.

\subsection{Adaptive Codebook-based Limited Feedback Protocol}
\label{ssec:protocol}


Motivated by the low overhead feedback requirement, we propose to exploit a \textit{codebook} structure for IRS control, where the BS sends only a quantized codeword index to the IRS.
Further, we consider 
\textit{adaptive} design of this codebook based on
channel variations.
We denote the instantaneous codebook as ${\mathcal C}[t] = \{ {\bf q}_m [t] \}_{m=1}^M $, where ${\bf q}_m [t] \in \mathbb{R}^{N_{\rm IRS}}$ is the $m$-th codeword (capacitance vector) and $M$ is the codebook size.
The codebook is stored and updated at the IRS through its control board (see Fig.~\ref{fig:system_merged}(a)).
We propose a novel \textit{limited feedback protocol}  consisting of four steps conducted per each coherence time block $t$, depicted in Fig.~\ref{fig:timeline}:

\begin{figure}[t]
  \includegraphics[width=.9\linewidth]{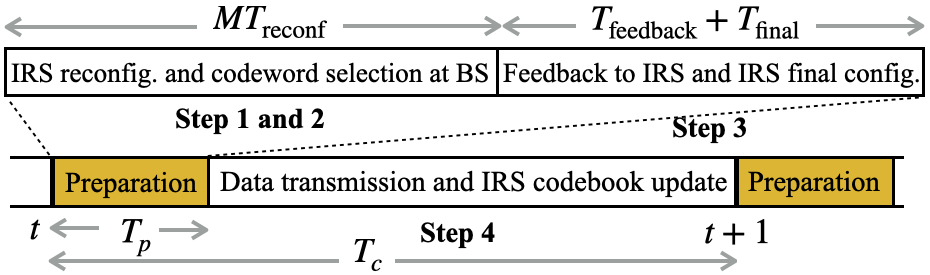}
  \centering
  \caption{Time frame structure of the proposed adaptive codebook-based limited feedback protocol for IRS-assisted communication.
  }
  \label{fig:timeline}
\end{figure}      


\begin{enumerate}[label={},leftmargin=0mm]
\item  {\it \textbf{Step 1. IRS channel sounding and reconfiguration}.}
While the UE transmits pilot symbols,
the IRS explores all of the $M$ capacitance vectors, i.e., ${\bf q}_m[t] \in \mathcal{C}[t]$, $m=1,...,M$.

\item {\it \textbf{Step 2. Codeword selection at BS.}}
The BS measures the effective channel ${\bf h}_{\rm eff}({\bf q}_m[t],t)$ and calculates the data-rate $ R({\bf q}_m[t],t)$  from \eqref{eq:obj:rate}, as IRS applies ${\bf q}_m[t]$, $m \negmedspace = \negmedspace 1,...,M$.
The BS obtains the codeword index 
\begin{equation}
    m^\star[t] \negmedspace = \negmedspace \underset{m \in \{1,...,M \}}{\arg\max} \; R({\bf q}_m[t],t).
\end{equation}

\item {\it \textbf{Step 3. Feedback to IRS and IRS final configuration.}} 
The BS feeds back $m^\star[t]$ to the IRS with $\lceil \log_2 M \rceil$ feedback bits. Then, the IRS tunes its meta-atoms with
 $ { \bf q}_{\star}[t] = {\bf q}_{m^\star[t]}[t] \in \mathcal{C}[t]$.

\item {\it \textbf{Step 4. Data transmission and IRS codebook update.}} The data transmission is conducted during the rest of the coherence time. During this period, the IRS obtains the next codebook $\mathcal{C}[t+1]$ either locally or with assistance from the BS.
\end{enumerate}


The benefits of our protocol include its
(i) simple procedure for IRS configuration, (ii) low-overhead feedback, and (iii)  adaptation to dynamic channels. 
%
Careful design of the codebook ${\mathcal C}[t]$ is critical to obtaining high data rates, since the codewords $\{{\bf q}_m[t]\}_{m=1}^M$ are the  solution candidates and the best one (${ \bf q}_{\star}[t]$ in \textit{\textbf{Step 3}}) among them is selected.\footnote{
$M$ should be limited due to the finite coherence time and non-negligible IRS reconfiguration time.
We consider that $M$ is predetermined in the protocol.}
We next propose two adaptive codebook design approaches for selecting $\mathcal{C}[t+1]$ in \textit{\textbf{Step 4}}, which utilize the previous IRS decisions and responses, with the understanding that the channels are in practice correlated between consecutive coherence times. 

%% file: method.tex
\section{Adaptive Codebook Design}
\label{sec:codebook}

For adaptive codebook design, we propose a 
low-overhead perturbation-based
approach in Sec.~\ref{ssec:RA}
and a deep neural network (DNN) policy-based approach in Sec.~\ref{ssec:DPIC}.
Then, we present a group control strategy, and quantify the time overhead and average data rate over one channel coherence block in Sec.~\ref{ssec:effective_rate}.

\subsection{Random Adjacency (RA) Approach}
\label{ssec:RA}

We first propose the {\it random adjacency} (RA) approach, which
can be viewed as a random perturbation-based method~\cite{mudumbai2007feasibility}
for codebook design.
%
Since the optimization~\eqref{eq:obj:rate}-\eqref{eq:con:rate} is conducted successively in time-correlated channels, the optimal solutions in adjacent time blocks are expected to be close to one another. 
The 
RA approach exploits this intuition by generating
multiple solution candidates 
around the previous solution.
The codebook resides and is updated at the IRS in this method.

Formally, the IRS obtains the codebook $\mathcal{C}[t+1] = \{{\bf q}_m[t+1]\}_{m=1}^M $, where  the $m$-th codeword  is updated by adding a random perturbation vector ${\bf z}_m[t] \in \mathbb{R}^{N_{\rm IRS}}$ to the previous solution ${\bf q}_\star[t]$ (obtained in \textit{\textbf{Step 4}} in Sec.~\ref{ssec:protocol}) as
\begin{equation}
    {\bf q}_m[t+1] = {\rm clip} ( {\bf q}_\star[t] + {\bf z}_m[t], [C_{\min}, C_{\max}] ), 
    \label{eq:RA:update}
\end{equation}
where ${\rm clip}(\cdot,[C_{\min}, C_{\max}])$ is an element-wise clip function ensuring constraint \eqref{eq:con:rate}.
Each entry of ${\bf z}_m[t]$ is generated from the uniform distribution $\mathcal{U}(-\delta, \delta)$, where $\delta$ is the maximum step size.
The RA approach follows \textit{\textbf{Step 1}}--\textit{\textbf{Step 4}}  in Sec.~\ref{ssec:protocol}, while in \textit{\textbf{Step 4}} the codewords  are updated by~\eqref{eq:RA:update}.


Intuitively, the RA approach becomes more effective as the number of codewords $M$ grows larger 
because more random points increase the chance of obtaining better codewords.
However, $M$ is limited due to the non-negligible IRS reconfiguration time and finite coherence time.
This makes the performance of the RA approach restricted due to the nature of the randomness and 
motives us to develop our next codebook update algorithm.

\subsection{DNN Policy-based IRS Control (DPIC) Approach}
\label{ssec:DPIC}


We next propose a DNN policy-based IRS control (DPIC) approach, aiming to learn \textit{policies} for updating the codewords.
In DPIC, the codebook
 resides at the IRS, as in RA.
However, the IRS now updates the codebook 
via information reception from the BS through the feedback link.
We consider that each codeword is updated \textit{independently} 
based on its prior deployments.
Henceforth, without loss of generality, we focus on the updates of $m$-th codeword.



\subsubsection{Low Overhead IRS Control via Direction Codebook}
To conduct the low overhead codeword update,
we introduce a fixed \textit{direction codebook} $\mathcal{D} = \{  {\bf d}_k \}_{k=1}^K$ 
where ${\bf d}_k \in \mathbb{R}^{N_{\rm IRS}}$, $k=1,...,K$.
The BS only transmits the index of a codeword in $\mathcal{D}$ 
to the IRS, which enables low feedback overhead for the codeword update.
We assume that  $\mathcal{D}$ is generated once at the beginning of the policy learning and shared at both the BS and IRS. 
The BS
employs a \textit{learning architecture} (discussed in Sec.~\ref{sssec:architecture})  to first obtain a continuous direction vector ${\boldsymbol u}_m[t] \in \mathbb{R}^{{N_{\rm IRS}}}$, from which it finds the index $k_m[t] \in \{ 1,...,K \}$ 
such that $k_m[t]$-th codeword ${\bf d}_{k_m[t]}$ in $\mathcal{D}$ has 
the  highest similarity to ${\boldsymbol u}_m[t]$.
The BS then feeds back the index $k_m[t]$ to the IRS, which the IRS uses to 
recover ${\bf d}_{k_m[t]}$
from $\mathcal{D}$, and
 updates the $m$-th codeword~as
\begin{equation}
    {\bf q}_m[t+1] = {\rm clip}({\bf q}_m[t] + {\bf d}_{k_m[t]}, [C_{\min}, C_{\max}]).
    \label{eq:DPIC:update}
\end{equation}

\subsubsection{IRS Control via Successive Decision Making}
\label{sssec:architecture}
The learning architecture at the BS consists of a DNN policy that determines ${\boldsymbol u}_m[t]$ and a quantization process that determines $k_m[t]$.
Our learning architecture consists of two phases: \textit{training phase} and \textit{utilization phase}. 
In the training phase, the BS aims to train the DNN policy to have an improved ${\boldsymbol u}_m[t]$ over time, while in the utilization phase the BS exploits the trained DNN policy without additional training.
In both phases, 
the BS first determines ${\boldsymbol u}_m[t]$ with the DNN policy based on the current information (i.e., the codeword ${\bf q}_m[t]$ in use and the effective channel ${\bf h}_{ {\rm eff} }({\bf q}_m[t], t )$).
Subsequently, the BS obtains $k_m[t]$ via a quantization process applied to  ${\boldsymbol u}_m[t]$ (described in Sec.~\ref{sssec:learning}\&\ref{sssec:utilization}).
The BS then feeds back $k_m[t]$ to the IRS, from which the IRS obtains the next codeword ${\bf q}_m[t+1]$ through \eqref{eq:DPIC:update}. 
The next codeword affects the subsequent information  at the BS (i.e., ${\bf q}_m[t+1]$ and ${\bf h}_{ {\rm eff} }({\bf q}_m[t+1], t+1 )$). 
The codeword update can thus be formulated as a
successive decision making process (Sec.~\ref{sssec:MDP}).

\subsubsection{MDP for Codeword Update}
\label{sssec:MDP}

We construct a Markov decision process (MDP) for the codeword update with the following state, action, and reward.

\textbf{State.} 
The state consists of 
information pertinent to the
environment evolution, which we define as
\begin{equation}
    {\bf s}_m[t] = \{
    {  {\bf h}_{ {\rm eff} }({\bf q}_m[t], t ) },
    {\bf q}_m[t]
    \}  \in \mathcal{S} =  \mathbb{R}^{2 N_{\rm BS} + {N_{\rm IRS}}},
    \label{eq:state}
\end{equation} 
where 
the real and imaginary parts of $ {\bf h}_{ {\rm eff} }({\bf q}_m[t], t )$ are stored as separate state dimensions.
%

\textbf{Action.} The action is the direction vector ${\boldsymbol u}_m[t]$:
\begin{equation}
    {\bf a}_m[t]
    = {\boldsymbol u}_m[t] 
    \in \mathcal{A}  =  [-\delta, \delta]^{N_{\rm IRS}},
    \label{eq:action}
\end{equation} 
where each entry of the action is bounded to the maximum step size, i.e., $[-\delta, \delta] \subset {\mathbb{R}}$. 
The action ${\bf a}_m[t]$ is used to determine the index $k_m[t]$
based on different processes in the training (Sec.~\ref{sssec:learning}) and utilization  (Sec.~\ref{sssec:utilization}) phases.
The next codeword ${\bf q}_m[t+1]$ is then obtained from $k_m[t]$  by \eqref{eq:DPIC:update}.

\textbf{Reward.}
The reward provides a measure of efficacy for policy learning.
We define the reward as
\begin{equation}
    r_m[t] =  R({\bf q}_m[t+1], t+1 )   
- \nu N_{{\rm clip},m}[t]
\in \mathbb{R},
\label{eq:DPIC:reward}
\end{equation}
where
$R({\bf q}_m[t+1], t+1 )$ denotes the data rate measured at time $t+1$ using codeword ${\bf q}_m[t+1]$, 
and $N_{{\rm clip},m}[t]$ denotes the number of  elements/dimensions in vector ${\bf q}_m[t] + {\bf d}_{k_m[t]}$ that hits the clipping threshold
in \eqref{eq:DPIC:update}.
$N_{{\rm clip},m}[t]$ is added as a  penalty  to 
avoid actions that result in the capacitance vectors violating~\eqref{eq:con:rate}.
In \eqref{eq:DPIC:reward}, $\nu >0$ is a weight parameter to match the order- of-magnitude of $R({\bf q}_m[t+1], t+1 )$ and $N_{{\rm clip},m}[t]$.

%
%
 %
 %



\subsubsection{Training Phase for DNN Policy Learning}
\label{sssec:learning}

We tailor a deep reinforcement learning (DRL) methodology to train the DNN policy with the formulated MDP.
We assume that the BS trains $M_{\rm A}$ different learning architectures, which are referred to as \textit{agents}. 
Agent $m\in \{1,...,M_{\rm A}\}$ has the DNN policy $\pi( {\bf s}_m[t]; {\bf w}_{\pi,m}): \mathcal{S} \rightarrow \mathcal{A}$
where ${\bf w}_{\pi,m}$ is the respective DNN weight parameters.
We consider that agent $m$ is trained with codeword $m$, $m\in \{1,...,M_{\rm A}\}$.

The \textit{actual action} of the agent $m$ (i.e., ${\bf d}_{k_m[t]}$ in \eqref{eq:DPIC:update}) is determined at the BS via the two following steps.
First, the BS adds a random noise vector ${\bf v}_m[t]$ to the output of the policy $\pi( {\bf s}_m[t]; {\bf w}_{\pi,m})$ to have more diverse responses and avoid getting trapped in local optima~\cite{sutton2018reinforcement}.
The BS uses the clip function to confine the output result to the feasible action space.
Second, the BS applies the {\it quantization process} using the direction codebook $\mathcal{D}$, through which
the BS determines the codeword index $k_m[t] \in \{1,...,K\}$ 
with closest Euclidean distance. In other words,
\begin{align}
    &k_m[t]  = 
    \nonumber
    \\
    &\underset{ k \in \{1,...,K\} }{\arg\min}  
    \| {\rm clip}(\pi( {\bf s}_m[t];\hspace{-.1mm}  {\bf w}_{\pi,m}) 
    + {\bf v}_m[t], \hspace{-.1mm} [-\delta,\delta])  - {\bf d}_k  \|_2.
    \label{eq:DPIC:behavior_policy}
\end{align}




To obtain $\pi( {\bf s}_m[t]; {\bf w}_{\pi,m})$,
we exploit the actor-critic network architecture of DRL.
This architecture consists of an
actor network $\pi( {\bf s}_m[t]; {\bf w}_{\pi,m})$ and a critic network $Q({\bf s}_m[t], {\bf a}_m[t]; {\bf w}_{Q,m})$ with DNN parameters ${\bf w}_{Q,m}$.
The actor selects an action using a policy, and the critic evaluates/criticizes the action to guide the actor network to take better actions.
However, using DNNs for reinforcement learning has been known to cause learning instability~\cite{sutton2018reinforcement}.
To stabilize the learning, 
we adopt the deep deterministic policy gradient (DDPG) approach~\cite{lillicrap2015continuous} for training the actor-critic networks.
The overall algorithm for training $M_{\rm A}$ agents in the limited feedback protocol is given in Algorithm~\ref{al:policy}. 
%
%

 \begin{algorithm}[h]
 \caption{Training $M_{\rm A}$ agents with actor-critic architecture}
 \label{al:policy}
 \begin{algorithmic}[1]
 \small
\State \textbf{Input.} 
$N_{\rm episode}$ (the number of learning episodes),
$N_{\rm timestep}$ (the duration of each episode), 
$C_{\min}$, $C_{\max}$, and $M_{\rm A}$.
  \State 
  Initialize 
  ${\bf w}_{Q,m}$ and ${\bf w}_{\pi,m}$.
   Empty the replay buffer $\mathcal{B}_m$, $m \in \{1,...,M_{\rm A}\}$. Set $\epsilon_{0} = (C_{\max}-C_{\min})/5$ and $\epsilon_{\min}=\epsilon_{0}/300$.  
  \For{$e=0,...,N_{\rm episode}-1$}
  \State \multiline{ Randomly generate
  $ \mathcal{C}[0] = \{ {\bf q}_m[0] \}_{m=1}^{M_{\rm A}}$.
  Update $\epsilon_e = \max \{ \epsilon_{\min}, 0.99 \epsilon_{e-1} \} $, if $e \ge 1$. }
  \For{$t=0,...,N_{\rm timestep}-1$}
    \State \multiline{ \textit{\textbf{Step 1. IRS channel sounding and reconfiguration.}} 
    The IRS meta-atoms are tuned following $\{{\bf q}_m[t]\}_{m=1}^{M_{\rm A}}$.
    }
    \State \multiline{ \textit{\textbf{Step 2. Inference at BS.}}
    Each agent $m \in \{1,...,M_{\rm A}\}$  computes $r_m[t-1]$ using \eqref{eq:DPIC:reward},
    forms ${\bf s}_m[t]$ in \eqref{eq:state}, and
     determines ${k_m[t]}$ using \eqref{eq:DPIC:behavior_policy} where ${\bf v}_m[t] \sim \mathcal{CN}({\bf 0}, \epsilon_e {\bf I}) $. \label{DPIC:tr:line:step2}
     }
    \State \multiline{ \textit{\textbf{Step 3. Feedback to IRS.}} The BS feeds back 
    $\{k_m[t]\}_{m=1}^{M_{\rm A}}$ to the IRS.
    }  
    \State \multiline{ \textit{\textbf{Step 4. IRS codebook update and BS training.}} 
    The IRS obtains $\mathcal{C}[t+1] = \{ {\bf q}_m[t+1] \}_{m=1}^{M_{\rm A}}$ by \eqref{eq:DPIC:update}.
    Each agent $m$ 
    stores $( {\bf s}_m[t-1], {\bf a}_m[t-1], r_m[t-1], {\bf s}_m[t] )$ in $\mathcal{B}_m$, samples  $\{({\bf s}_i, {\bf a}_i, {r}_i, {\bf s}'_{i}) \}_{i=1}^{N_{\rm batch}}$ from $\mathcal{B}_m$,
    and
    updates ${\bf w}_{Q,m}$ and ${\bf w}_{\pi,m}$ via the DDPG algorithm in \cite{lillicrap2015continuous}.
     }
     \EndFor
     \EndFor
 \end{algorithmic}
 \end{algorithm}

\subsubsection{Utilization Phase with Trained DNN Policies}
\label{sssec:utilization}

In the utilization phase, we utilize the $M_{\rm A}$ trained agents to conduct codebook updates without additional training of the agents.
The BS partitions $M$ codewords among $M_{\rm A}$ agents. 
We consider that codeword $m \in \{1,...,M\}$ is allocated to agent $j[m] = \mod(m-1,M_{\rm A})+1 \in \{1,...,M_{\rm A}\}$.
If $M_{\rm A}=1$, a single agent handles all $M$ codeword updates, which we refer to as a single-agent DPIC (SDPIC).
If $M_{\rm A}>1$, multiple agents handle the $M$ codeword updates, which we refer to as multi-agent DPIC (MDPIC).
Utilizing more agents often improves  
performance due to the ensemble learning principle~\cite{goodfellow2016deep}. 

The DPIC approach follows \textit{\textbf{Step 1}}--\textit{\textbf{Step 4}} of the protocol in Sec.~\ref{ssec:protocol}. Additionally, in \textit{\textbf{Step 2}}, the BS constructs 
${\bf s}_m[t] = \{ {\bf h}_{\rm eff}({\bf q}_m[t],t), {\bf q}_m[t] \}$ and determines $ {k_m[t]}$ in \eqref{eq:DPIC:behavior_policy} by using ${\bf v}_m[t]=0$ (no random noise addition) and ${\bf w}_{\pi,j[m]}$ (instead of ${\bf w}_{\pi,m}$). 
In \textit{\textbf{Step 3}}, 
the BS additionally feeds back $\{ k_m[t] \}_{m =1}^M$ to the IRS for codebook updates, which incurs a total of $\lceil \log_2 M \rceil + M \lceil \log_2 K \rceil $ feedback bits.
In \textit{\textbf{Step 4}}, the BS also updates the codebook.

 \subsection{Group Control, Time Overhead, and Effective Data Rate}
\label{ssec:effective_rate}

We consider a \textit{group control}~\cite{yang2020intelligent}, where IRS meta-atoms are partitioned into multiple groups and the same capacitance is applied for the meta-atoms belonging to the same group. This reduces the dimension of the design variables, and thus
the BS can conduct the training/inference in a timely manner.
%
We thus neglect the computation time overhead
 of our methods, and
define the {\it time overhead} $T_p$ 
shown in Fig. \ref{fig:timeline} as
\begin{equation}
    T_p = M  T_{\rm reconf} + T_{\rm feedback} + T_{\rm final}.
    \label{eq:time_overhead}
\end{equation}
In \eqref{eq:time_overhead}, $M T_{\rm reconf}$ denotes the total time for $M$ IRS reconfiguration (in \textit{\textbf{Step 1}} in Sec.~\ref{ssec:protocol}) used in both RA and DPIC approaches, where $T_{\rm reconf}$ is the time for each IRS reconfiguration, and $T_{\rm feedback}$ denotes the time required for the feedback from the BS to the IRS (in \textit{\textbf{Step 3}}), which  is different for the RA and DPIC approach.
For the RA approach, the feedback time is
\begin{equation}
    T_{\rm feedback} \negmedspace = \negmedspace \frac{ \lceil \log_2 M \rceil} {W R_{\rm feedback}},
\end{equation}
where 
 $R_{\rm feedback}$ (bits/s/Hz) is the unit data rate for the feedback link. 
For the DPIC approach, the feedback time is
\begin{equation}
    T_{\rm feedback} = \frac{\lceil \log_2 M \rceil + M \lceil \log_2 K \rceil} {W R_{\rm feedback}}.
\end{equation}
%
Lastly, $T_{\rm final}$ denotes the execution time of the final IRS reconfiguration (in \textit{\textbf{Step 3}}).
If the selected index $m^\star[t]$ coincides with the last configuration in \textit{\textbf{Step 1}}, the IRS 
does not need to change the configuration, i.e., $T_{\rm final} \negmedspace = \negmedspace0 $; otherwise $T_{\rm final} \negmedspace=\negmedspace T_{\rm reconf}$. 


To measure the average data rate during one coherence block
with coherence time $T_c$,
we define
{\it effective data rate} as
\begin{equation}
    R_{\rm eff}[t] = \frac{T_c-T_p}{T_c} W \log_2
    \bigg(1+ \frac{P  \| {\bf h}_{\rm eff}( {\bf q}_\star[t], t) \|_2^2 }
    { \sigma^2 } \bigg),
    \label{eq:effrate}
\end{equation}
where $T_c - T_p$ is the actual data transmission time and ${\bf q}_\star[t] \in \mathcal{C}[t]$ is the selected codeword for the IRS configuration.
The above metric captures the tradeoff between the data rate and the time overhead $T_p$.
As $M$ increases, the data rate 
may increase due to having larger number of reconfigurations, while $T_p$ also increases leading to decreasing of $T_c - T_p$.
In Sec. \ref{sec:sim}, we evaluate the data rate and effective data rate under different~$M$.


%% file: sim.tex
\section{Numerical Evaluation and Discussion}
\label{sec:sim}

In this section, we describe the simulation setup in Sec.~\ref{ssec:sim:setup}, and then
present and discuss the simulation results in Sec. \ref{ssec:sim:indoor}.

\subsection{Simulation Setup}
\label{ssec:sim:setup}


To emulate practical IRS reflection behavior,
we recover 
$\Gamma(C, \theta) $ 
from
the data in Fig.~4 and Table~1 of \cite{chen2020angle}, where the ranges of $C$ and $\theta$ are $(C_{\min} , C_{\max}) = (0.4 , 2.7)$ pF and $(0^o, 90^o)$, respectively.
We set $f = 5.195$ GHz and consider only azimuth coordinates as in~\cite{chen2020angle}.
We assume $T_c=5$ ms, $N_{\rm BS} = 5$ and $N_{\rm IRS} = 200$.
We consider a group control with the number of groups $N_{\rm G} = 10$, where 
$N_{\rm IRS}/ N_{\rm G}= 20$ meta-atoms are controlled by each common capacitance.
The BS antenna spacing is $d_{\rm BS}=\lambda/2$, and the IRS meta-atom spacing is $d_{\rm IRS} = \lambda/10$ where $\lambda = c/f$ and $c=3 \times 10^8$ m/s.
%
The BS and IRS are 
located at $ (0,0)$ m and $(90,30)$ m, respectively. 
The initial UE position 
is randomly generated within the circle with radius $5$ m at $(100,0)$ m.
The UE is moving with the velocity $v = 3$ km/h and constant azimuth angle  generated
from $\mathcal{U}(0,2\pi)$.
%
We set $P = 20$ dBm,  $\sigma^2 = -80$ dBm, $W=10$ MHz, $R_{\rm feedback}=0.1$ bits/s/Hz, and
$T_{\rm reconf} = 100 \mu s$ \cite{abadal2020programmable}.\footnote{The reconfiguration time of IRS is determined by the characteristics of the control board and the internal communication between the control board and the meta-surface. 
The reconfiguration speed is typically a few kHz \cite{abadal2020programmable}.} 

For RA, we set $\delta = (C_{\max}-C_{\min})/5$.
For DPIC, we set 
$N_{\rm batch} = 32$, $\nu = W$, and  $|\mathcal{B}_m| = 5\times10^{5}$, $m \in \{1,...,M_{\rm A}\}$.
For the DNNs, we consider a fully connected neural network with two hidden layers consisting of $400$ and $300$ neurons, respectively, with ReLU activation function. 
%
For the DNN policy, 
in the output layer the tanh function is employed, and the output is scaled by $\delta = (C_{\max} - C_{\min})/4$ to bound the actions. 
We set $|\mathcal{D}|=K=2048$, where each codeword in $\mathcal{D}$ is constructed by RVQ~\cite{au2007performance} ranging within $[-\delta, \delta]^{N_{\rm G}}$.
We employ the Adam optimizer for training.
For MDP, we normalize the values as
$ {\bf h}_{\rm eff}(\cdot) \leftarrow \sqrt{{P}/({\sigma^2 N_{\rm BS} N_{\rm G} }) } \times {\bf h}_{\rm eff}(\cdot) $ and ${\bf q}_m \leftarrow 10^{12} \times {\bf q}_m $ in \eqref{eq:state}, ${\bf a}_m \leftarrow 10^{13} \times {\bf a}_m $ in \eqref{eq:action}, and $R(\cdot) \leftarrow R(\cdot)/W$ in~\eqref{eq:DPIC:reward}.

%
We adopt a multi-path geometric channel model~\cite{tse2005fundamentals} for the IRS-BS, UE-BS, and UE-IRS channels. 
We model (i) the IRS-BS channel as Rician channel with $K$-factor of $5$~\cite{tse2005fundamentals}, $10$ non line-of-sight (NLoS) signal paths, and a path loss exponent (PLE) of $2$,
(ii) the UE-BS channel with only NLoS signals of $10$ paths and a PLE of $3.75$,
and (iii) the UE-IRS channel with only NLoS signals of $10$ paths and a PLE of $2.2$.
%
%
The small scale fading factors evolve according to a first-order Gauss-Markov process \cite{sklar2001digital} with a time correlation coefficient of $0.95$ (corresponding to speed of $3$ km/h of the UE/scatterers), and the angle of each path varies over every coherence time by an amount generated from $\mathcal{U}(-0.1^o,0.1^o)$.


\subsection{Simulation Results and Discussion}
\label{ssec:sim:indoor}

\begin{figure}[t]
%
\centering
\begin{subfigure}{.9\linewidth}
  \centering
  \includegraphics[width=\linewidth]{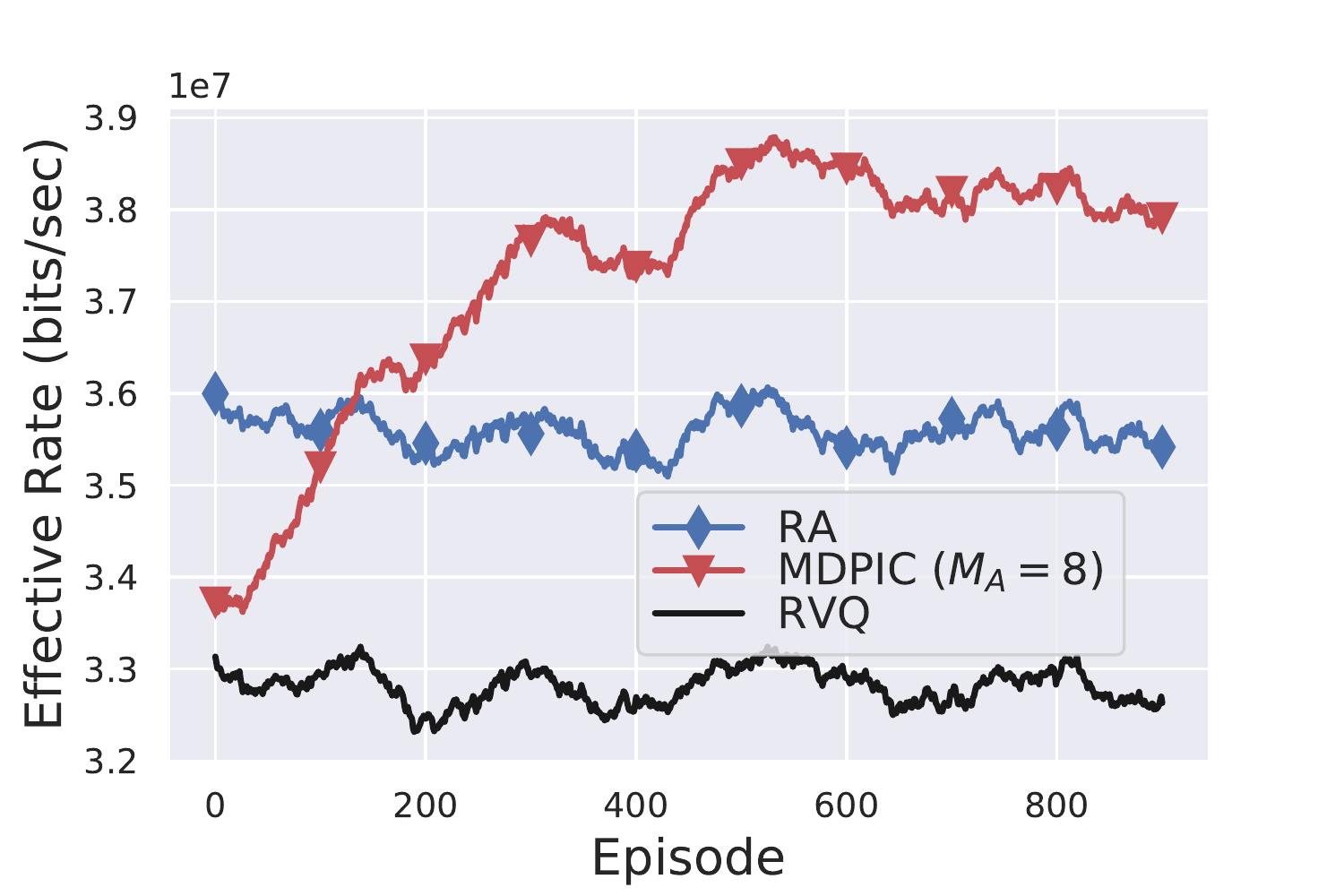}
  \caption{ Effective data rate  along episodes
  }
  \label{fig:train}
\end{subfigure}
\begin{subfigure}{.9\linewidth}
  \centering
  \includegraphics[width=\linewidth]{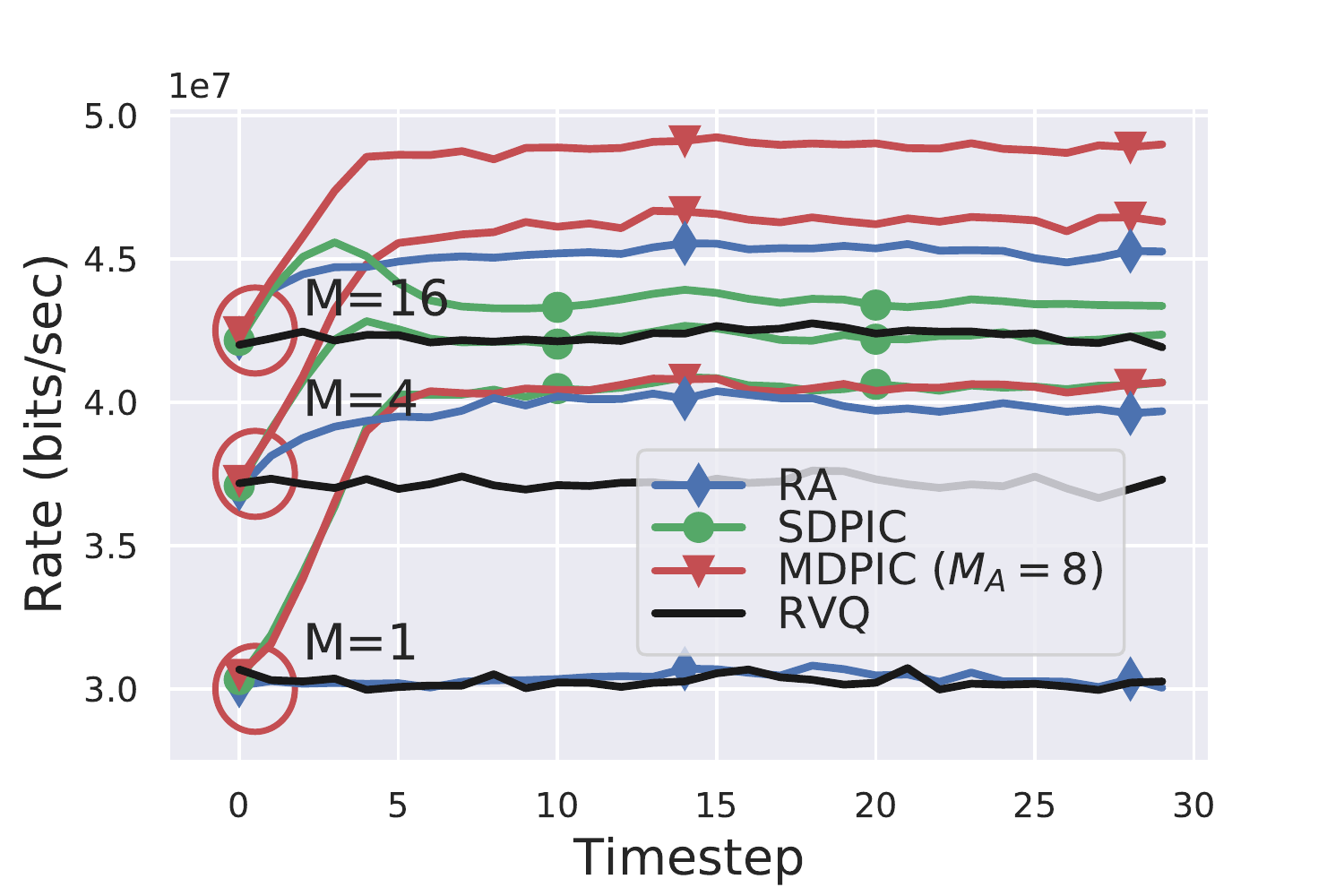}
 \vspace{-6mm}
  \caption{ Data rate along timesteps  
  }
  \label{fig:rate_util}
\end{subfigure}
\begin{subfigure}{.9\linewidth}
  \centering
  \includegraphics[width=\linewidth]{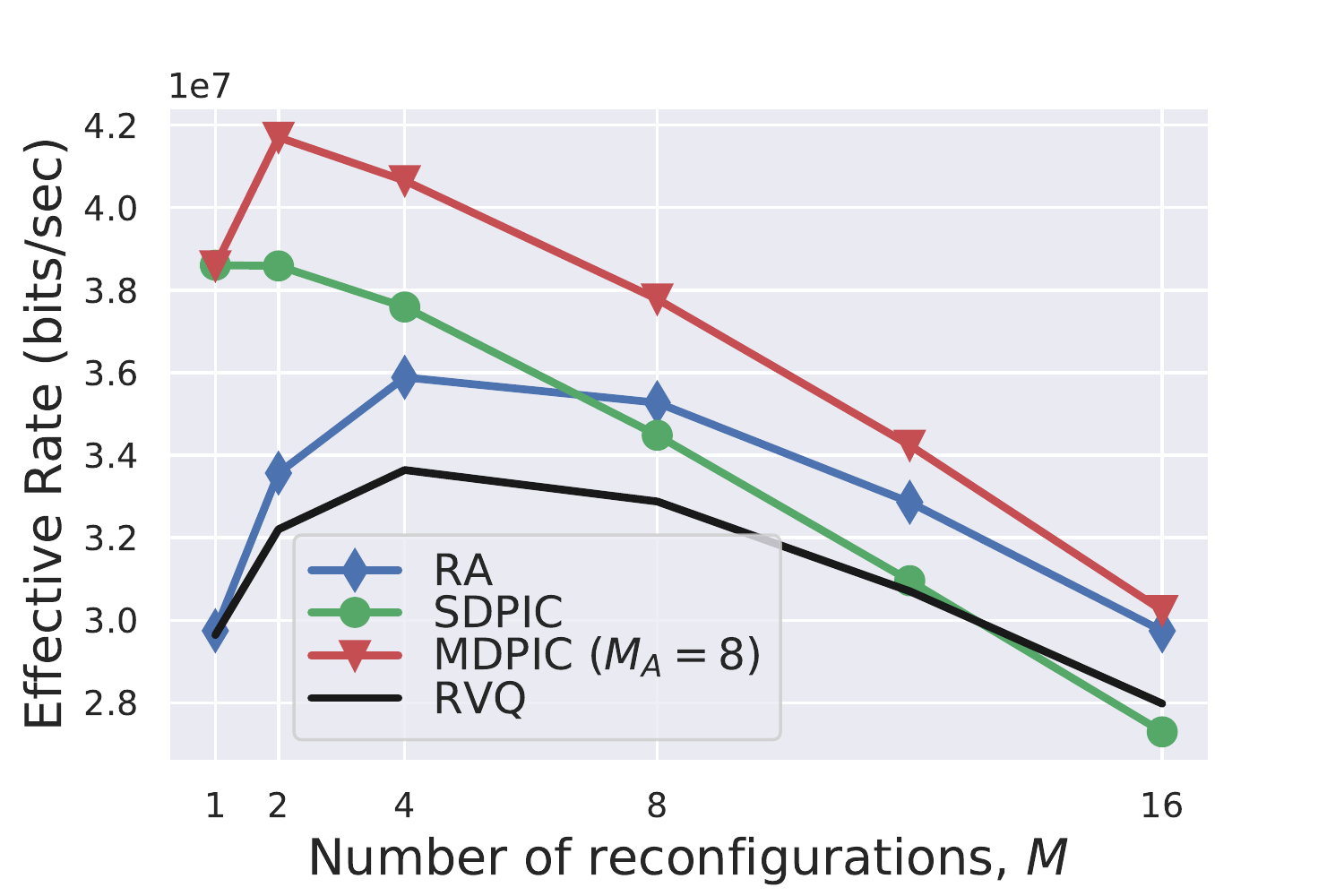}
  \caption{ Effective data rate along $M$ 
  }
  \label{fig:eff_util}
\end{subfigure}
\caption{Performance evaluation of our methodology. (\subref{fig:train}) corresponds to the training phase, while
(\subref{fig:rate_util})-(\subref{fig:eff_util}) correspond to the utilization phase.
}
\label{fig:NLoS}
\end{figure}

%
We train $M_{\rm A}=8$ agents in the training phase with $1000$ episodes, each with $500$ timesteps (coherence blocks). 
Each episode has different realizations of the UE-IRS channel, UE-BS channel, IRS-BS channel, initial UE location, and UE moving direction.
Our baseline is the RVQ codebook~design \cite{au2007performance}.
Fig. \ref{fig:NLoS}(\subref{fig:train})
shows the effective data rate averaged over the timesteps. Each data point is a moving average over the previous 100 episodes.
The performance of MDPIC is improved over time since the agents (specifically, the DNN policies) are trained to conduct better codebook updates.

We then evaluate our proposed algorithms in the utilization phase with $2000$ episodes each with $30$ timesteps.
Fig. \ref{fig:NLoS}(\subref{fig:rate_util}) depicts the average data rate over the utilization episodes.
Our proposed schemes -- RA, SPDIC, and MDPIC -- update the codebook adaptively 
by using previous observations (i.e., previously used codeword and end-to-end channel) to improve the data rate over time.
Our methods obtain their peak performance 
within $4$-$5$ timesteps.
Overall, as the number of IRS reconfiguration $M$ increases, a higher data rate is achieved.
The MDPIC yields better data rate compared to those of the SDPIC and RA due to the advantage of using multiple agents.

Fig. \ref{fig:NLoS}(\subref{fig:eff_util}) shows the  effective data rate
along $M$.
The effective data rate in \eqref{eq:effrate} captures the tradeoff between the data rate and the time overhead discussed in Sec.~\ref{ssec:effective_rate}.
The MDPIC method shows the best performance in terms of effective data rate for any $M$, and
has the highest value at $M^\star=2$. 
As $M$ grows larger than $2$, the increased time overhead outweighs the improvement of the data rate, leading to the decrease of the effective data rate. 



%% file: conc.tex
\section{Conclusion}
\label{sec:conc}

In this paper, we introduced a novel signal model incorporating the practical IRS reflection behavior.
To address the design challenges faced with IRS control under
multi-path dynamic channels and low-overhead feedback requirements,
we proposed the codebook-based limited feedback protocol.
We proposed two adaptive codebook designs: the random adjacency  and the deep neural network policy-based IRS control.
Through simulations, we showed that the data rate and effective data rate performances are improved by the proposed schemes.